# Auto-locking Waveguide Amplifier System for Lidar and Magnetometric Applications

A. Pouliot, H.C. Beica, A. Carew, A. Vorozcovs, G. Carlse and A. Kumarakrishnan
Department of Physics and Astronomy, York University, Toronto Ontario M3J 1P3

## ABSTRACT

We describe a compact waveguide amplifier system that is suitable for optically pumping rubidium magnetometers. The system consists of an auto-locking vacuum-sealed external cavity diode laser, a semiconductor tapered amplifier and a pulsing unit based on an acousto-optic modulator. The diode laser utilises optical feedback from an interference filter to narrow the linewidth of an inexpensive laser diode to ~500 kHz. This output is scannable over an 8 GHz range (at 780 nm) and can be locked without human intervention to any spectral marker in an expandable library of reference spectra, using the autolocking controller. The tapered amplifier amplifies the output from 50 mW up to 2 W with negligible distortions in the spectral quality. The system can operate at visible and near infrared wavelengths with MHz repetition rates. We demonstrate optical pumping of rubidium vapour with this system for magnetometric applications. The magnetometer detects the differential absorption of two orthogonally polarized components of a linearly polarized probe laser following optical pumping by a circularly polarized pump laser. The differential absorption signal is studied for a range of pulse lengths, pulse amplitudes and DC magnetic fields. Our results suggest that this laser system is suitable for optically pumping spin-exchange free magnetometers.

**Keywords:** Auto-locked diode lasers, magnetometers, laser spectroscopy, tapered amplifiers

## 1. INTRODUCTION

There is widespread interest in developing a new generation of spin-exchange free rubidium magnetometers[1] for geophysical exploration. These magnetometers can be used in airborne surveys for the improved detection of metal and mineral deposits. Magnetometers require laser sources for optical pumping. The laser source should be capable of being frequency stabilized with respect to rubidium atomic transitions without the need for human intervention. Additionally, the laser pulses should have pulse widths of about 100 ns, repetition rates of several kilohertz and power outputs of a few Watts. Here we describe a pulsed laser system based on auto-locked diode laser systems (ALDLS)[2-5] that is capable of addressing all the required specifications.

During the last forty years, there have been significant improvements in the sensitivity of vapour cell magnetometers used for the rapid detection of small magnetic fields in airborne and terrestrial surveys. These surveys are capable of detecting magnetic anomalies and magnetized rocks associated with deposits of metals and minerals. Vapour cell magnetometers rely on optically pumping alkali vapours using high power light sources.

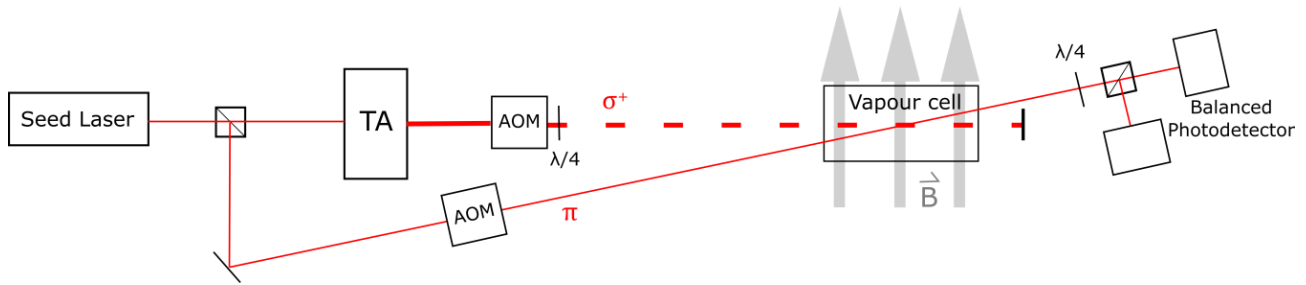

Figure 1: Schematic of the rubidium magnetometer consisting of an auto-locked seed laser and tapered amplifier. Separate acousto-optic modulators (AOMs) are used to generate an amplified, pulsed, pump laser that is circularly polarized and a continuous wave, linear polarized probe laser with the same optical frequency. The pump and probe are aligned at an angle of a few milliradian through a rubidium vapour cell. A balanced detector and a current amplifier is used to record the differential absorption of the probe laser in the presence of a magnetic field.

Figure 1 shows the implementation of a rubidium magnetometer based on references 6 and 7. A circularly polarized pump laser (carrying angular momentum σ+) is used to optically pump alkali vapour into a single magnetic ground state sublevel. Figure 2 shows a simplified level diagram for a D1 transition.

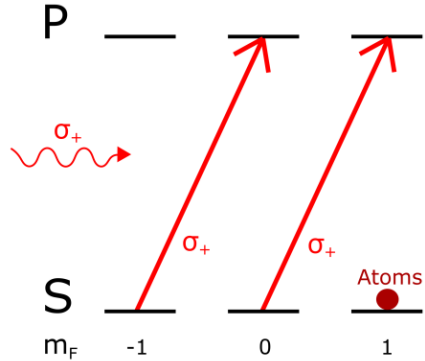

Figure 2: Simplified level diagram for an alkali D1 transition. Atoms are optically pumped by circularly polarized light into the extreme magnetic sublevel of the ground state where they can no longer be addressed by the laser.

The vapour becomes spin polarized and transparent to the pump laser when the optical pumping is complete. Therefore, monitoring the transmission of the pump laser serves as a convenient method of probing the atomic polarization due to optical pumping. An external magnetic field drives a periodic change in population between adjacent ground state magnetic sublevels at the Larmor frequency. As a result, the orthogonal, circularly polarized components of the probe laser are differentially absorbed and exhibit Larmor oscillations. A balanced photodetector can be used to record these Larmor oscillations and measure the magnetic field.

In a vapour cell, the timescale for the decay of the differential absorption signal is dominated by the transit time of optically pumped atoms. This timescale can be significantly extended by the addition of a buffer gas so that the motion of atoms is dominated by diffusion. Alternatively, wall coatings can be used to greatly reduce depolarization due to wall collisions so that the signal decay time is independent of transit time. Under these conditions the dominant depolarizing effect is radiation trapping. This effect can be reduced by introducing a small concentration of a quenching gas which ensures that excited atoms return to the ground state via collisions before the spontaneous emission of a photon. Depolarization can still occur under these conditions through spin-exchange collisions between ground state rubidium atoms. Such collisions preserve the angular momentum of the atoms and exchange their hyperfine states. After the collision, depolarization occurs due to the opposite direction of spin precession for differing hyperfine states. In spin-exchange free magnetometers[1], the frequency of rubidium-rubidium collisions is increased so that depolarization is avoided by constantly cycling the atoms between the two hyperfine states and preventing evolution into other magnetic sublevels, thereby allowing sensitivities of ~1 fT Hz$^{-1/2\,\mathrm{s}}$.

Alkali magnetometers have been realized using both D1 and D2 transitions (795 nm and 780 nm respectively for rubidium). In a vapour cell without buffer gas, the effect of optical pumping is similar for both D1 and D2 transitions, with the most extreme magnetic sublevel being populated by a circularly polarized pump. The differential absorption signal arises due to differential absorption of the two orthogonal, circularly polarized components of the probe laser, this signal is larger on the D1 transition. In the presence of a buffer gas, the excited state is collisionally broadened and de-excitation due to collisions or spontaneous emission populate all magnetic sublevels of the ground state with equal probability. Under these conditions the atomic polarization for the D1 transition is unaffected, but it is reduced by a factor of two for the D2 transition. These considerations make the D1 transition preferable for magnetometric experiments.

## 2. DESCRIPTION OF LASER SYSTEM

We have developed a unique, low cost, vacuum-sealed, auto-locked external cavity diode laser system[2-5]. ALDLS units are integrated using components from original equipment manufacturers (OEM) coupled with specially machined parts and powerful central processors. The laser's master oscillator depends on optical feedback from a narrow band interference filter to realize a narrow laser line width (~500 kHz)[8,9]. The thermally stabilized laser cavity can be evacuated within minutes and vacuum-sealed for several months making the system insensitive to environmental temperature and pressure fluctuations. The optical feedback from the interference filter can be adjusted from outside the cavity using a vacuum feedthrough. The ALDLS can be locked or scanned with respect to a spectral line without the need for human intervention using a digital controller that is capable of storing a variety of algorithms in its memory for laser frequency stabilization using techniques such as pattern matching and first or third derivative feedback. The laser cavity relies on an interchangeable optics kit consisting of a laser diode and optical feedback elements to operate in the desired wavelength range. Therefore, the master oscillator can be designed to optically pump both the D2 (780 nm) and D1 (795 nm) absorption lines in rubidium. The laser source can also address spectral lines associated with both $^{85}$Rb and $^{87}$Rb isotopes. The ALDLS technology enables additional features such as power amplification of the master oscillator's output of 100 mW to several Watts using semiconductor waveguides[10] as well as rapid amplitude modulation using acousto-optic modulators (AOMs) and radio frequency (RF) electronics. A schematic diagram of the pulsed laser system is shown in Figure 3.

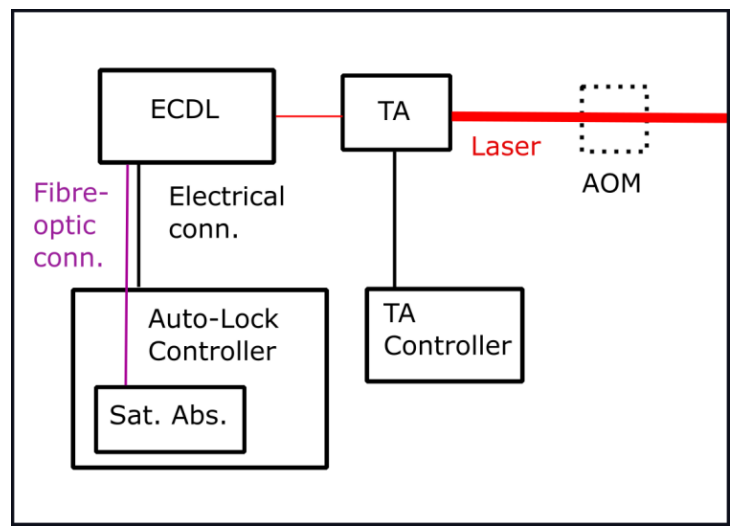

Figure 3: Schematic of ALDLS system consisting of an external cavity diode laser (ECDL) with an auto-lock controller and waveguide tapered amplifier (TA). Amplitude modulation is achieved by pulsing an acousto-optic modulator (AOM) with RF circuits.

The output of the master oscillator is fibre coupled through a beam splitter into an auto-lock controller containing a saturation-absorption spectrometer and control electronics for frequency stabilization with respect to Rb spectra. Another output of the beam splitter is aligned through a 2-W semiconductor waveguide amplifier to increase the power output. The AOM utilizes TTL switches to produce suitably short pulses for optical pumping.

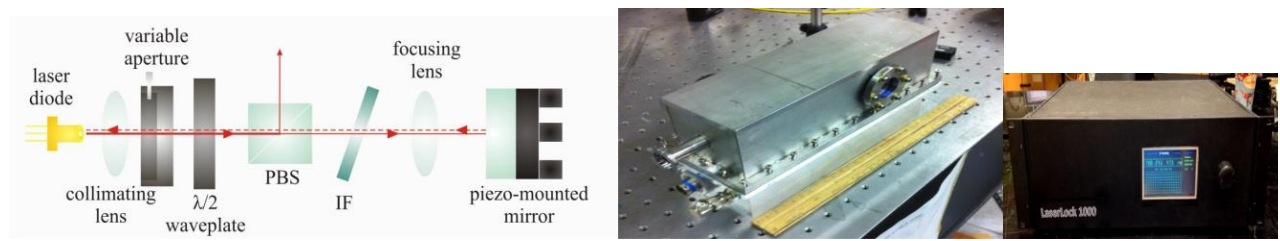

Figure 4: Schematic showing components of laser head, and pictures of the laser head and auto-lock controller.

Figure 4 shows a schematic diagram of the laser head, a laser head, and an auto-lock controller. Optical feedback from a narrowband interference filter[8,9] is used to realize a laser linewidth of 500 kHz. Laser operation has been demonstrated at 780 nm and 633 nm (see Figure 5) based on rubidium and iodine spectroscopy, respectively. We characterized the system performance through measurements of the Allan deviation of the beat note between two lasers (Figure 5), and through a measurement of the Allan deviation of the lock stability of a single laser. The laser frequency was stabilized using different auto-locking algorithms selected from the digital controller's library for these studies. We used third derivative feedback for iodine spectroscopy, and both pattern matching and first derivative feedback for rubidium spectroscopy. The laser linewidth and lock stability allowed precision measurements of gravitational acceleration with an accuracy of 3 parts-per-billion (ppb) using a state-of-the-art industrial sensor. Our studies also showed that the correction signals were reduced by nearly an order of magnitude by evacuating the air in the laser cavity. Under laboratory conditions, the Allan deviation of the beat note between two identical lasers was measured to be $2.5 \times 10^{-11}$ for a measurement time $\tau = 40$ s (Figure 5). The Allan deviation of the lock stability of a single laser is $2 \times 10^{-11}$ for $\tau = 80$ s, which suggests a similar level of performance. These specifications compare favourably with respect to different types of diode lasers[11-13].

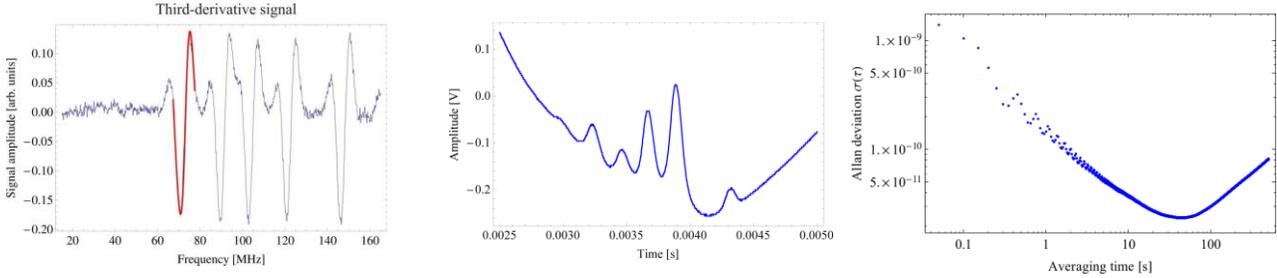

Figure 5: Third derivative error signals in iodine at 633 nm, Rb spectra at 780 nm, and Allan Deviation of 30 MHz beat note between two 780 nm lasers showing a minimum value of $2.5 \times 10^{-11}$ for $\tau = 40$ s

The pulsed laser system relies on the amplification of highly monochromatic light from an auto-locking diode laser system (ALDLS) using a tapered, semiconductor waveguide amplifier. The amplifier consists of two sections: a waveguide, and a tapered gain region. The tapered geometry of the gain region couples the amplified light into several spatial modes. This allows the amplifier to operate at high currents without increasing the energy density of the beam inside the device to the point where non-linear effects can cause catastrophic self-focusing. Spatially filtered laser light from a monochromatic source is coupled into the waveguide section and diffracts into the tapered section such that the tapered gain region is completely filled, allowing maximum amplification of the seed light[10]. Figure 6 shows the performance of a 2 W tapered amplifier seeded by 17 mW of light from a seed laser locked to a Rb spectral line

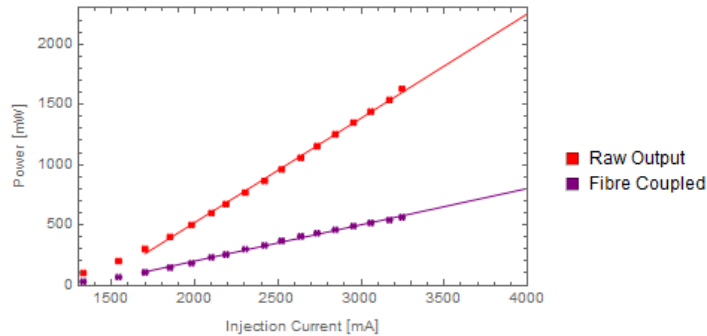

Figure 6: Output power of 2-Watt tapered amplifier as a function of injection current recorded with and without fibre coupling at a temperature of 16°C. The power input from the seed laser was 17 mW. The straight lines show the extrapolated power output for the maximum allowed injection current.

The amplifier introduces two detrimental effects in the seed light. Firstly, the spatial profile of the beam is distorted by the large number of spatial modes associated with the amplifier cavity. Seeding with a filtered TEM00 mode limits the output to a subset of spatial modes, but the beam is still far from Gaussian, assuming a more "flat topped" profile with a few local maxima. Secondly, the spectral profile of the output beam is modified by the broadband amplified spontaneous emission (ASE) from the TA cavity. The simplest way to reduce these effects is to spatially filter the output. This process eliminates undesired spatial modes and a large portion of the ASE as well[14]. However, it comes at the cost of the output power of the beam as shown in Figure 6.

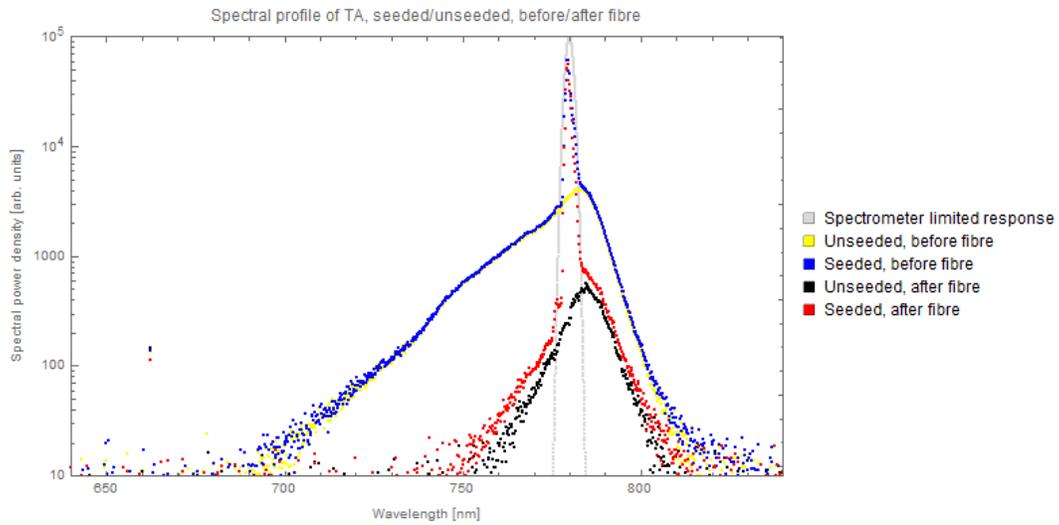

Figure 7: Spectral characterization of a 2-Watt waveguide amplifier operating at 780 nm with a fibre coupled spectrometer. The full width at half maximum (FWHM) of spectral response function of the spectrometer (1 nm) is shown in grey. This FWHM is much larger than the linewidth (500 kHz) of the seed laser. The spectrometer transmission is recorded before and after an optical fibre that is used to spatially filter the output of the amplifier. At each location, the ASE is recorded by blocking the seed light from the ALDLS. The spectrum of the amplified light is also recorded before and after the fibre.

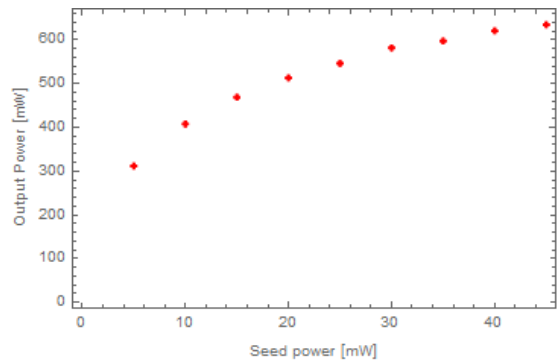

Figure 8: Waveguide amplifier output as a function of seed-power. This data was recorded with a 2-Watt amplifier driven with an injection current of 1980 mA at a temperature of 16°C.

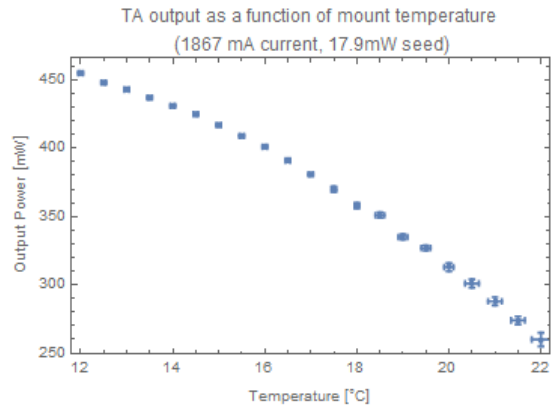

Figure 9: Output power of 2-Watt waveguide amplifier as a function of temperature.

We have characterized the relative intensity of the ASE and its reduction through fibre coupling by using a spectrometer with a resolution of 1 nm as shown in Figure 7. The ASE consists of partially coherent light spanning several nanometers. We find that the intensity of ASE is three to four orders of magnitude smaller than the amplified light at the seed wavelength.

The output power of the amplifier has been measured as a function of the power of the seed laser as shown in Figure 8. The data shows the saturation of the power output of the amplifier as the seed power approaches 30 mW. These results also suggest that a single seed laser (power output of 100 mW) can drive up to three waveguide amplifiers.

## 3. MAGNETOMETRIC STUDIES

In this section we present preliminary results of a rubidium magnetometer. Demonstrating the effect of optical pumping using the ALDLS. We use the magnetometric configuration shown in figure 1 based on references 6 and 7. Ideally the laser would address the rubidium D1 transition at 795 nm to maximize the atomic polarization and the differential absorption signal. Since the ALDLS is designed to operate at 780 nm, our experiments are carried out on the rubidium D2 transition.

The pump and probe laser are derived from the same seed laser, and the pump is amplified by a tapered amplifier. The two beams are overlapped at an angle of approximately 10 milliradian in a 5 cm rubidium cell at room temperature that contains no buffer gas. Identical 80 MHz AOMs are deployed in each beam for amplitude modulation and for ensuring that both lasers interact with the same velocity class. Two elliptical magnetic fields coils are used to apply a DC field perpendicular to the direction of laser propagation.

The differential absorptions signal is measured using a balanced detector that contains two reverse-biased photodiodes with 10 ns risetimes. The signal is amplified by a current amplifier with a gain of 1 microamp per volt and recorded using an oscilloscope with a bandwidth of 100MHz. The data was obtained while operating on the F=3→F'=4 transition in $^{85}$Rb.

Figure 10 shows two typical datasets, obtained in the Earth's ambient field for two different durations of the optical pulsing pulse. The timescale for the signal decay is consistent with transit time for rubidium atoms at room temperature. The longer pulse results in a 3-fold increase in the signal amplitude due to increased optical pumping but there is a decrease in the contrast of Larmor oscillations. For this dataset, the largest signal (a further 3-fold enhancement) was obtained with pulse powers of 230 mW, and pulse lengths of 3 µs. This data can be used to develop optical pumping simulations.

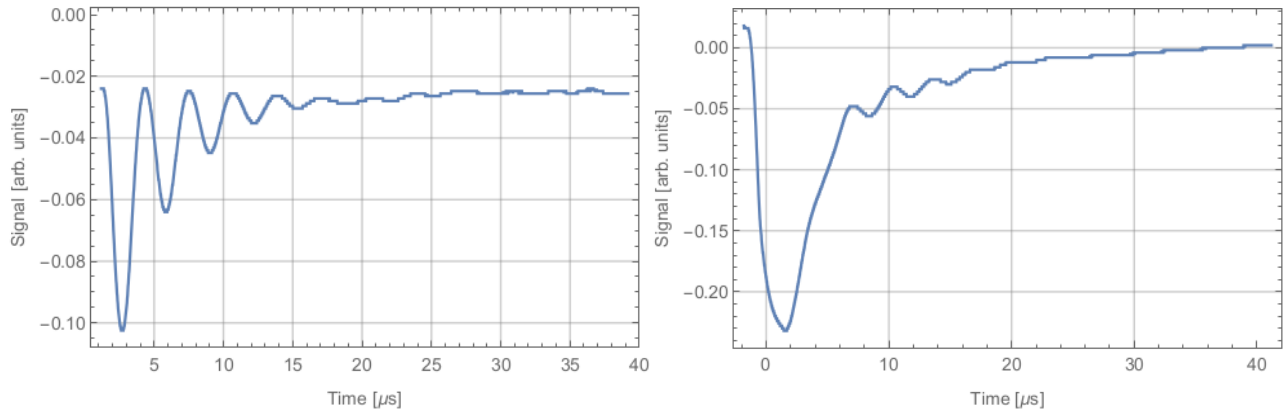

Figure 10: Effect of optical pumping in the Earth's ambient magnetic field acquired with a pulse power of 164.1 mW, and a probe power of 71.8 µW. The trace on the left displays the signal due to a 100 ns optical pumping pulse and the image on the right was recorded with a 3 µs pulse.

Figure 11 shows the effect of varying the external magnetic field due to the coils. The precession frequency is extracted from a fit to the Fourier transform of each signal. The fit is the sum of a Gaussian and an exponential decay. The variation in the extracted Larmor frequency shows an offset from zero-frequency. We attribute this effect to the component of the Earth's magnetic field perpendicular to both the applied field and the direction of propagation. The Larmor frequency shows a minimum at a nonzero value of the applied field due to the component of the Earth's field along the axis of the coils. This data gives a slope of 0.59 ± 0.02 MHz/G whereas the expected slope defined by the Zeeman shift in $^{85}$Rb is 0.47 MHz/G. We attribute this difference to the inhomogeneity of the magnetic field across the length of the cell and the error in the calibration of the magnetic field sensor.

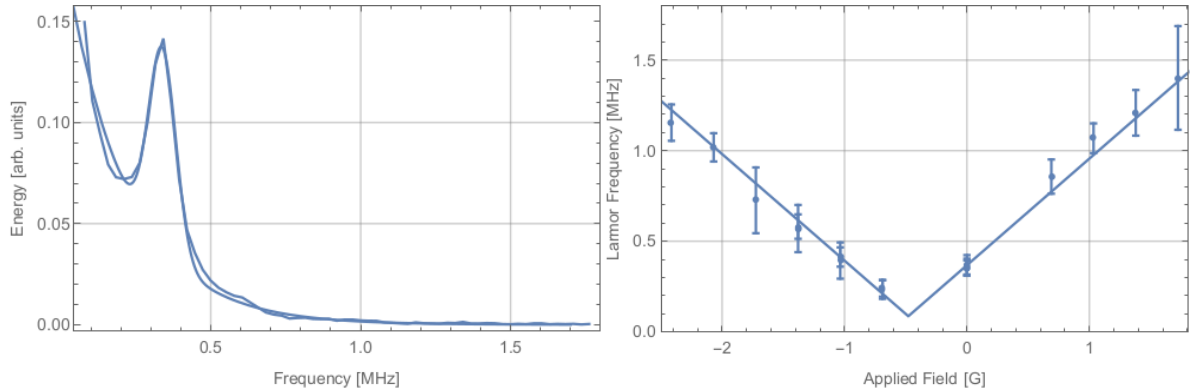

Figure 11: The left panel is the Fourier transform of a signal used to extract the frequency of Larmor oscillations. The right panel shows the oscillation frequency as a function of the magnetic field due to the coils. Here the pump power was 230 mW, the pump power was 71.8 µW.

## 4. CONCLUSIONS

Our results suggest that compact ALDLS based on tapered amplifiers can be used for optically pumping rubidium magnetometers. We plan to extend our studies to vapour cells with both buffer and quenching gasses so that we can explore optical pumping in the spin-exchange free regime. These results will be compared against multi-level optical pumping simulations.